\documentstyle[11pt,newpasp,twoside,epsf]{article}
\def\edcomment#1{\iffalse\marginpar{\raggedright\sl#1\/}\else\relax\fi}
\reversemarginpar

\begin{document}
\title{A Numerical Study of the Chemo-Dynamical Evolution of Elliptical
 Galaxies}
 \author{Daisuke Kawata and Brad K.\ Gibson}
\affil{ Centre for Astrophysics and Supercomputing,
Swinburne University of Technology, Hawthorn VIC 3122, Australia}

\begin{abstract}
We investigate the chemo-dynamical evolution
of elliptical galaxies, to understand the origin of
the mass-dependence of photometric properties such as the 
colour-magnitude relation (CMR). Our three-dimensional TREE
N-body/SPH numerical simulation takes into account both Type II
and Type Ia supernovae and follows the evolution of the abundances
of several chemical elements.
We derive the photometric properties of the simulation
end-products and compare them with the observed CMR.
\end{abstract}

\section{Introduction}

 The colour-magnitude relation (CMR) of elliptical galaxies is 
a natural byproduct of galaxy formation scenarios involving galactic wind.
 However, those scenarios were initially based upon
a simple model which ignored the internal structure and complex
star formation history inherent to galaxy formation.
Numerical simulations are a powerful complement which allows one to 
treat complex physical processes in galaxy formation self-consistently.
We have developed a code that can calculate
the dynamical, chemical, and photometric evolution 
of galaxies self-consistently.
To understand further the physics of elliptical
galaxy formation, we examine whether realistic numerical simulations
can reproduce the observed CMR.

\section{Methods and Results}

 Our simulations model galaxy formation as an evolution of 
a low-spin top-hat over-dense sphere, namely a seed galaxy
(Kawata 2001). 
Since we are interested in the evolution of a seed galaxy with
different masses, we simulate representative $L^*$ and sub-$L^*$
seed galaxies with masses of $4\times10^{12} M_\odot$ and $2\times10^{11}
M_\odot$, respectively. The dynamics of collisionless dark matter and stars
is calculated using a Tree N-body code,
and the gas component is modeled using SPH.
The code includes radiative cooling, star formation, 
supernova feedback, and metal enrichment. We take into account
both Type II (SNe II) and Type Ia (SNe Ia) supernovae (SNe),
and chemical enrichment from intermediate mass stars.
We assume that SNe feedback is released as 
thermal ($E_{\rm th}$) and kinetic ($E_{\rm kin}$)
energy, and their ratio is given by a parameter
$f_v=E_{\rm kin}/(E_{\rm th}+E_{\rm kin})$.
This parameter controls the magnitude of the effect of SNe. 
To explore this effect, we carry out simulations with strong ($f_v=0.04$)
and weak feedback ($f_v=0.005$).
Figure 1 shows the comparison of the $V-K$ CMR for simulation 
end-products and Coma cluster galaxies (Bower et al. 1992).
These results are obtained by coupling to the population synthesis
package of KA97 (Kodama \& Arimoto 1997).
The slope of the strong feedback model is steeper and
in better agreement with the observed slope than that of the weak feedback
model. Figure 1 also shows the metallicities for the models and 
demonstrates that the CMR is driven by metallicity effects.
The history of SNe II, which roughly traces the star formation history,
shows that star formation stops abruptly at an early epoch
(right panels in Fig.~1).
This cessation of star formation is caused by the galactic wind.
The galactic wind suppresses further chemical enrichment and leads
to bluer colours especially in the low mass system.
Moreover, as seen in the CMR, the strong feedback
enhances the mass dependence of the suppression of chemical enrichment
and causes the steeper slope in the CMR.
We find that the strong effect of SNe feedback
is required to explain the observed CMR.
We also found that SNe Ia play a crucial role in driving and 
maintaining a galactic wind, and therefore in contributing to
the evolution of elliptical galaxies.

\begin{figure}
\plotone{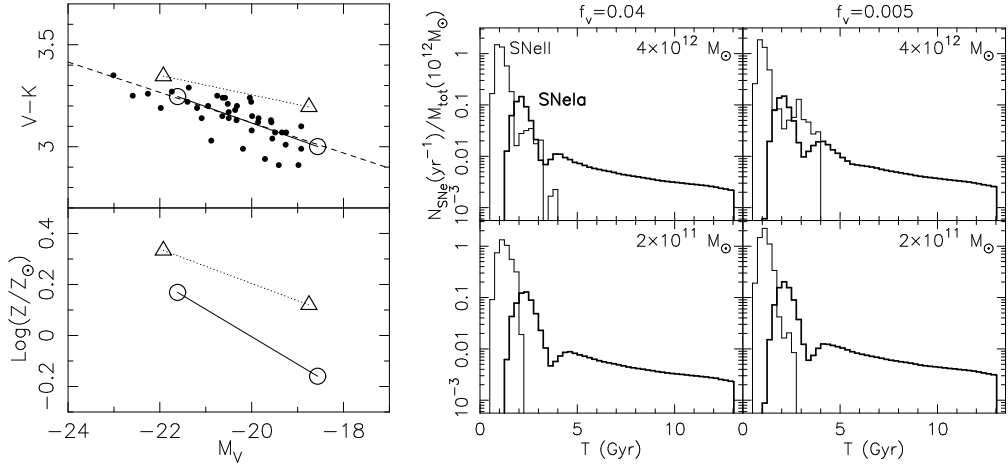}
\caption{ Upper left panel: comparison of the CMRs for the
simulation end-products and the Coma cluster galaxies (small dots).
Lower left panel: the metallicities against the absolute $V$
band magnitude. The triangles (circles) connected by dotted (solid) lines
indicate the results of models with $f_v=0.04$ ($f_v=0.005$).
Right panels: time variations of the event rate of SNe II (thin lines)
and SNe Ia (thick lines) for all the models.
}
\end{figure}


\begin{references}
\reference 
 Bower, R.G., Lucey, J.R., \& Ellis R.S.\ 1992, \mnras, 254, 589
\reference
 Kawata, D.\ 2001, \apj, 558, 598
\reference
 Kodama, T., \& Arimoto, N.\ 1997, \aap, 320, 41
\end{references}
\end{document}